\documentclass[floatfix,showpacs,superscriptaddress,prl,twocolumn]{revtex4}
\usepackage{amsbsy,amssymb,amsmath,bm}
\usepackage{graphicx,color}
\usepackage{amsfonts}
\usepackage{amssymb}
\usepackage{amsmath}

\input{tcilatex}

\begin{document}

\title{Dynamics of the particle - hole pair creation in graphene}
\author{M. Lewkowicz}
\affiliation{\textit{Applied Physics Department, Ariel University Center of Samaria,
Ariel 40700, Israel}}
\author{B. Rosenstein}
\email{vortexbar@yahoo.com}
\affiliation{\textit{Electrophysics Department, National Chiao Tung University,Hsinchu
30050,} \textit{Taiwan, R. O. C}.}
\date{\today }

\begin{abstract}
The process of coherent creation of particle - hole excitations by an
electric field in graphene is quantitatively described. We calculate the
evolution of current density, number of pairs and energy\ after switching on
the electric field. In particular, it leads to a dynamical visualization of
the universal finite resistivity without dissipation in pure graphene. We
show that the DC conductivity of pure graphene is $\frac{\pi }{2}\frac{e^{2}%
}{h}$rather than the often cited value of $\frac{4}{\pi }\frac{e^{2}}{h}$.
This value coincides with the AC conductivity calculated and measured
recently at optical frequencies. The effect of temperature and random
chemical potential (charge puddles) are considered and explain the recent
experiment on suspended graphene. A possibility of Bloch oscillations is
discussed within the tight binding model.
\end{abstract}

\maketitle

1. \textit{Introduction. }It has been demonstrated recently that a graphene
sheet, especially one suspended on leads, is one of the purest electronic
systems. Electronic mobility reaches values of $2\cdot 10^{5}cm^{2}/\left(
Vs\right) $ and might be yet improved \cite{GeimPRL08,Andrei08} indicating
that transport in samples of submicron length is most likely ballistic. In a
simplified model of a single graphene sheet (neglecting scattering processes
and electron interactions) the chemical potential is located right between
the valence and conductance bands and the Fermi "surface" consists of two
Dirac points of the Brillouin zone \cite{Castro}. A lot of effort has been
devoted to the question of transport in pure graphene due to the surprising
fact that the DC conductivity is finite without any dissipation process
present. A widely accepted value of the "minimal conductivity" at zero
temperature, 
\begin{equation}
\sigma _{1}=\frac{4}{\pi }\frac{e^{2}}{h},  \label{sigma1}
\end{equation}%
was calculated very early on using the Kubo formula in a simplified Dirac
model as well as in the tight binding model \cite%
{Fradkin,Ando02,Castro,Katsnelson06}. Within this approach one starts with
the AC conductivity and takes a zero frequency limit typically with certain
"regularizations" (like finite temperature, disorder strength $\eta $ etc.)
made and removed at the end of the calculation. As noted by Ziegler \cite%
{Ziegler06} the order of limits makes a difference and several other values
different from $\sigma _{1}$ were provided for the \textit{same} system. The
standard value $\sigma _{1}$ is obtained using a rather unorthodox procedure
when the DC limit $\omega \rightarrow 0$ is made \textit{before }the zero
disorder strength limit $\eta \rightarrow 0$ is taken. If the order of
limits is reversed one obtains \cite{Ziegler06}

\begin{equation}
\sigma _{2}=\frac{\pi }{2}\frac{e^{2}}{h}.  \label{sigma2}
\end{equation}%
When the limit is taken holding $\omega =\eta $ one can even obtain a value
of $\sigma _{3}=\pi \frac{e^{2}}{h}$ \cite{Ziegler06}, thus solving the
"missing $\pi $" problem. Indeed, at least early experiments on graphene
sheets on Si substrates provided values roughly 3 times larger than $\sigma
_{1}$ \cite{Novoselov05}. Recent experiments on suspended graphene \cite%
{Andrei08} demonstrated that the DC conductivity is lower, $1.7\sigma _{1}$,
as temperature is reduced to $4K$. Hence one still faces the question of
what is the proper theoretical value. Since the conductivity of clean
graphene in the infinite sample is a well defined physical quantity there
cannot be any ambiguity as to its value.

In contrast both the experimental and the theoretical situation for the AC
conductivity in the high frequency limit is quite different. The
theoretically predicted value in the Dirac model is $\sigma _{2}$
independent of frequency under condition $\omega >>T/\hbar $ \cite%
{Ando02,Varlamov07}. The Dirac model becomes inapplicable when $\omega \ $is
of order of $\gamma /\hbar =4\cdot 10^{15}Hz$ or larger, where $\gamma
=2.7eV $ is the hopping energy of graphene. It was shown theoretically\
using the tight binding model and experimentally in \cite{GeimScience08}
that the optical conductivity at frequencies higher or of order $\gamma
/\hbar $ becomes slightly larger than $\sigma _{2}$. Moreover, in light
transmittance measurements at frequencies down to $2.5\cdot 10^{15}Hz$\ it
was found equal to $\sigma _{2}$ within 4\%. The model does not contain any
other time scale capable of changing the limiting value of AC conductivity
all the way to $\omega ->0$. Therefore one would expect that the DC
conductivity even at zero temperature is $\sigma _{2}$ rather than $\sigma
_{1}$. As we show in this note this is indeed the case.

The basic physical effect of the electric field is a coherent creation of
electron - hole pairs mostly near Dirac points. To effectively describe this
process we develop a dynamical approach to charge transport in clean
graphene using the "first quantized" approach to pair creation physics
similar to that used in relativistic physics \cite{Gitman}. To better
visualize the phenomenon of resistivity without dissipation, we describe an
experimental situation as closely as possible by calculating directly the
time evolution of the electric current after switching on an electric field.
In this way the use of a rather formal Kubo or Landauer formalism is avoided
and as a result no regularizations are needed. The effects of temperature
and of charge fluctuations or "puddles" are investigated and explain the
temperature dependence of conductivity\ measured recently in suspended\
graphene \cite{Andrei08}. Although we consider an infinite sample the
dynamical approach allows us to obtain qualitative results for finite
samples by introducing time cutoffs like ballistic flight time. Various
other factors determining transport can be conveniently characterized by
time scales like the relaxation time for scattering of phonons or impurities.

\textit{2. Time evolution of the current density at zero temperature. }%
Electrons in graphene are described sufficiently well by the 2D tight
binding model of nearest neighbour interactions \cite{Castro}, namely with
(second quantized) Hamiltonian being sum over all the links connecting sites
on two triangular sublattices $A,B$. The Hamiltonian in momentum space is

\begin{equation}
\widehat{H}=\int_{BZ}%
\begin{pmatrix}
c_{A}^{\dag } & c_{B}^{\dag }%
\end{pmatrix}%
H%
\begin{pmatrix}
c_{A} \\ 
c_{B}%
\end{pmatrix}%
;\text{ \ \ }H=%
\begin{pmatrix}
0 & h \\ 
h^{\ast } & 0%
\end{pmatrix}%
,  \label{H}
\end{equation}%
where $h\left( k\right) =-\gamma \sum_{\alpha }e^{i\mathbf{k}\cdot \delta
_{\alpha }}$ with $\gamma $ being the hopping energy; $\delta _{1}=\frac{a}{3%
}\left( 0,\sqrt{3}\right) ,$ $\ \delta _{2,3}=\frac{a}{3}\left( \pm \frac{3}{%
2},-\frac{\sqrt{3}}{2}\right) $\ are the locations of nearest neighbours on
the honeycomb lattice separated by distance $a\simeq 3\mathring{A}$. In the
Brillouin zone of the lattice there are two Dirac points $K_{-}=\frac{2\pi }{%
a}\left( \frac{1}{3},\frac{1}{\sqrt{3}}\right) ,\ K_{+}=\frac{2\pi }{a}%
\left( \frac{2}{3},0\right) $ in which the energy gap between the valence
and the conduction band vanishes. Expansion around $K_{-}$, $h\left(
k\right) =\hslash v_{g}\exp \left( -i\frac{\pi }{3}\right) \left( \Delta
k_{x}+i\Delta k_{y}\right) ,$ where the graphene velocity is $v_{g}=\frac{%
\sqrt{3}}{2}\frac{a\gamma }{\hslash },$ leads to relativistic equations for
the Weyl field constructed as $\psi _{1}=\psi _{1}^{W},$ $\ \psi _{2}=e^{-i%
\frac{\pi }{3}}\psi _{2}^{W}$.

Let us first consider the system in a constant and homogeneous electric
field along the $y$ direction switched on at $t=0.$ It is described by the
minimal substitution $\mathbf{p}=\hslash \mathbf{k}+\frac{e}{c}\mathbf{A}$
with vector potential (choosing a gauge in which the scalar potential is
zero) $\mathbf{A}=\left( 0,-cEt\right) $. Since the crucial physical effect
of the field is a coherent creation of electron - hole pairs mostly near
Dirac points a convenient formalism to describe the pair creation is the
"first quantized" formulation described in detail in \cite{Gitman}. The
second quantized state at $T=0$ which evolves from the zero field state in
which all the negative energy ($-\left\vert h\left( k\right) \right\vert $)
states are occupied is uniquely characterized by the first quantized
amplitude $\psi _{k}\left( t\right) =%
\begin{pmatrix}
\psi _{k1}\left( t\right)  \\ 
\psi _{k2}\left( t\right) 
\end{pmatrix}%
$ obeying the matrix Schroedinger equation $i\hslash \partial _{t}\psi
=H\psi $ in sublattice space with the initial condition 
\begin{equation}
\psi _{k}\left( t=0\right) =u_{k};\text{ \ }u_{k}=\frac{1}{\sqrt{2}}%
\begin{pmatrix}
1 \\ 
-h^{\ast }/\left\vert h\right\vert 
\end{pmatrix}%
.  \label{u}
\end{equation}%
Here $u_{k}$ is found as a negative energy solution of the time independent
Schroedinger equation prior to switching on the electric field, $%
Hu_{k}=-\left\vert h\right\vert u_{k}$.

A physical quantity is usually conveniently written in terms of $\psi $. For
example the current density (multiplied by factor $2$ due to spin) is

\begin{equation}
J_{y}=-2e\int_{BZ}\psi _{k}^{\dag }\left( t\right) \frac{\partial H\left( 
\mathbf{p}\right) }{\partial p_{y}}\psi _{k}\left( t\right)  \label{J}
\end{equation}%
To first order in electric field $\psi _{k}=e^{\frac{i}{\hslash }\left\vert
h\right\vert t}\left( u_{k}+E\xi _{k}+...\right) $ and consequently $%
J_{y}=J_{0}+E\sigma ,$ where

\begin{eqnarray}
J_{0} &=&-\frac{2e}{\hslash }\int_{BZ}u_{k}^{\dag }\frac{\partial H\left( 
\mathbf{k}\right) }{\partial k_{y}}u_{k}=\frac{2e}{\hslash }\int_{BZ}\frac{%
\partial \left\vert h\right\vert }{\partial k_{y}};  \label{cond} \\
\sigma \left( t\right) &=&\int_{BZ}\sigma _{k}\left( t\right) ;  \notag \\
\sigma _{k}\left( t\right) &=&-\frac{2e}{\hslash }\left[ u_{k}^{\dag }\frac{%
\partial H\left( \mathbf{k}\right) }{\partial k_{y}}\xi _{k}+\xi _{k}^{\dag }%
\frac{\partial H\left( \mathbf{k}\right) }{\partial k_{y}}u_{k}-\frac{e}{%
\hslash }tu_{k}^{\dag }\frac{\partial ^{2}H\left( \mathbf{k}\right) }{%
\partial k_{y}^{2}}u_{k}\right] .  \notag
\end{eqnarray}%
The solution of the Schroedinger equation for the correction $\xi _{k}$ is

\begin{eqnarray}
\xi _{k} &=&\frac{ie}{2\hslash ^{2}}t^{2}\frac{\partial \left\vert
h\right\vert }{\partial k_{y}}u_{k}  \label{corr} \\
&&+\frac{ie}{8\left\vert h\right\vert ^{3}}\left( h^{\ast }\frac{\partial h}{%
\partial k_{y}}-cc\right) \left( 1-e^{-2i\left\vert h\right\vert t/\hslash }-%
\frac{2i\left\vert h\right\vert }{\hslash }t\right) v_{k},  \notag
\end{eqnarray}%
where $v_{k}=\frac{1}{\sqrt{2}}%
\begin{pmatrix}
1 \\ 
h^{\ast }/\left\vert h\right\vert%
\end{pmatrix}%
$. Substituting this into Eq.(\ref{cond}) the conductivity becomes

\begin{equation}
\sigma _{k}\left( t\right) =\frac{e^{2}}{\hslash }\left[ -\frac{\partial
^{2}\left\vert h\right\vert }{\partial k_{y}^{2}}\frac{2t}{\hslash }-\frac{1%
}{4\left\vert h\right\vert ^{4}}\left( h^{\ast }\frac{\partial h}{\partial
k_{y}}-cc\right) ^{2}\sin \left( \frac{2\left\vert h\right\vert }{\hslash }%
t\right) \right] .  \label{sig_k}
\end{equation}%
The zero field current $J_{0}$ and the first term (linear in time) in the
conductivity vanish upon integration over the Brillouin zone, since one can
choose it to be $\int_{BZ}=\int_{-\pi /a}^{\pi /a}dk_{x}\int_{-2\pi /\left(
3^{1/2}a\right) }^{2\pi /\left( 3^{1/2}a\right) }dk_{y}$ and the integrand
is a derivative of a periodic function.

The integral of the second part (oscillatory in time) of $\sigma _{k}\left(
t\right) $ gives the result shown in Fig.1. After an initial increase over
the natural time scale $t_{\gamma }\equiv \hslash /\gamma =2.5\cdot
10^{-16}s $, it approaches $\sigma _{2}$, Eq.(\ref{sigma2}), via
oscillations. The amplitude of oscillations decays as a power 
\begin{equation}
\frac{\sigma }{\sigma _{2}}=1+\frac{\sin \left( 2t/t_{\gamma }\right) }{%
2t/t_{\gamma }}
\end{equation}
for $t>>t_{\gamma }$. The limiting value is dominated by contributions from
the vicinity of the two Dirac points in the integral of Eq.(\ref{cond}). The
contribution of a Dirac point is obtained for $t>>t_{\gamma }$ by
integrating to infinity (in polar coordinates centered at the Dirac point)

\begin{equation}
\frac{\sigma }{2}=\frac{e^{2}}{\hslash }\frac{1}{\left( 2\pi \right) ^{2}}%
\int_{\varphi =0}^{2\pi }\int_{q=0}^{\infty }\sin \left( \varphi \right) ^{2}%
\frac{\sin \left( 2v_{g}qt\right) }{q}=\frac{e^{2}}{h}\frac{\pi }{4},
\label{Dpointssigma}
\end{equation}%
does not influence the result.

\begin{figure}[tbp]
\includegraphics[width=8.5cm]{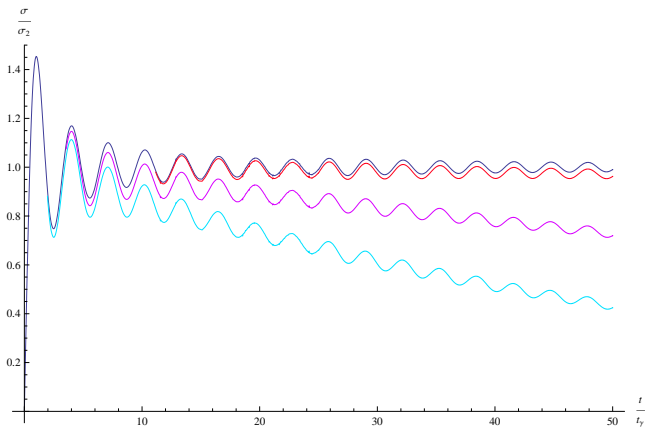}
\caption{Evolution of the current density $\protect\sigma \left( t\right)
=J\left( t\right) /E$ after a DC electric field is switched on at $t=0$.
Unit of time is $t_{\protect\gamma }=\hbar /\protect\gamma $. Conductivity
is compared to its "dynamical" value $\protect\sigma _{2}=\frac{\protect\pi 
}{2}\frac{e^{2}}{h}$. The zero temperature conductivity (blue) approaches $%
\protect\sigma _{2}$, while finite temperature depresses the pair creation
and eventually the current density vanishes as $\frac{1}{tT}$.}
\label{Fig. 1}
\end{figure}

A physical picture of this resistivity without dissipation is as follows.
The electric field creates electron - hole excitations in the vicinity of
the Dirac points in which excitations are massless relativistic fermions.
For such particles the absolute value of the velocity is $v_{g}$ and cannot
be altered by the electric field and is not related to the wave vector $%
\mathbf{k}$\textbf{. }On the other hand, the orientation of the velocity is
influenced by the applied field.\textbf{\ }The electric current is $e\mathbf{%
v}$, thus depending on orientation, so that its projection on the field
direction $y$ is increased by the field. The energy of the system
(calculated in a way similar to the current) is increasing continuously if
no channel for dissipation is included. Obviously at some time the system
goes beyond linear response into Bloch oscillations which are briefly
discussed below. We have performed a similar calculation for the evolution
of the current density for an AC electric field switched on at $t=0$. After
a short transient one obtains the value of the DC conductivity $\sigma _{2}$
independent of frequency. This is consistent with both the Kubo formula
derivations \cite{Varlamov07} and optical experiments \cite{GeimScience08}.

\textit{3. The temperature dependence and effect of charge "puddles". }At
finite temperature $T$ within the first quantized formalism one adds the
contributions of all the energies including positive ones weighted with the
Boltzmann factor. Due to electron - hole symmetry the contribution to
conductivity of a positive energy state with momentum $\mathbf{k}$ is minus
that of the contribution of the negative energy state with the same wave
vector. This results in the thermal factor 
\begin{equation}
\sigma _{T}\left( t\right) =\int_{BZ}\tanh \left( \frac{\left\vert
h\right\vert }{T}\right) \sigma _{k}\left( t\right) .  \label{sigT}
\end{equation}

The first term still vanishes, while the second gives a depressed value
compared to that at $T=0$, see Fig. 1. Moreover, the conductivity vanishes
at the large time limit. This is easy to appreciate qualitatively: the
contributions from the vicinity of the Dirac points, $\left\vert
h\right\vert <<T$, which were the main contributors to $\sigma \left(
T\right) $ are effectively suppressed. Physically this suppression can be
understood as follows. As mentioned above the finite resistivity of pure
graphene is due to pair creation by an electric field near Dirac points. The
pair creation is maximal when in the initial state the valence band is full
and the conductance band is empty. Thermal fluctuations create pairs as
well. In the formalism we adopted the finite temperature initial state is
described by the density matrix which specified the number of incoherent
pairs present in the energy range near the Dirac points. Therefore pair
creation by an electric field is less intensive due to the diminished phase
space available and the conductivity vanishes at large times.

Under assumption of Dirac point dominance, $T<<\gamma $ (definitely covering
the temperature range $T<200K$ beyond which scattering is not negligible 
\cite{GeimPRL08}), the expression can be simplified in the same way as Eq.(%
\ref{Dpointssigma}), 
\begin{equation}
\sigma _{T}\left( t\right) =\frac{e^{2}}{h}\int_{q=0}^{\infty }\tanh \left( 
\frac{\hbar v_{g}q}{T}\right) \frac{\sin \left( 2v_{g}tq\right) }{q}\text{,}
\label{sigT_appr}
\end{equation}%
and is a monotonically decreasing function of the product $tT$. For $%
t>>t_{\gamma ,}$ $\sigma _{T}\left( t\right) =\frac{e^{2}}{h}\frac{\hbar }{tT%
}$.

\begin{figure}[tbp]
\includegraphics[width=8.5cm]{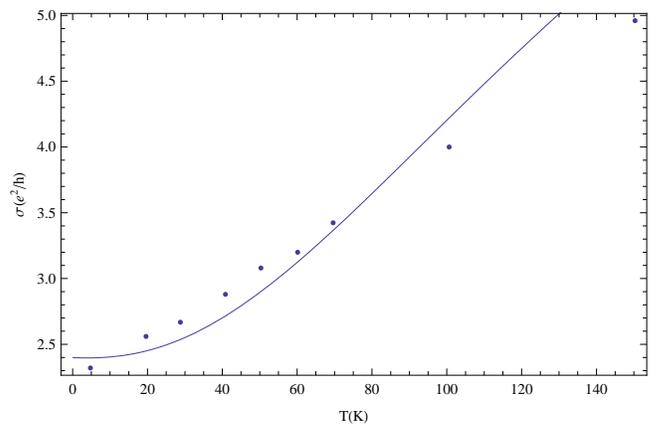}
\caption{The minimal conductivity as a function of temperature for time $%
t_{bal}=500t_{\protect\gamma }$ is compared with measured in the $0.5\protect%
\mu m$ long sample in ref.\protect\cite{Andrei08}. Values for the random
Fermi energy are also taken from ref.\protect\cite{Andrei08}.}
\label{Fig. 2}
\end{figure}

Assuming ballistic transport in a finite sample of submicron length
determining an effective ballistic time $t_{b}\,$, this contribution cannot
explain the \textit{increase} of conductivity with temperature in suspended
graphene reported in \cite{Andrei08}. However, there is an important source
of positive contribution to conductivity even in the ballistic regime. It
was clearly demonstrated that a sample close to minimal conductivity
consists of positively and negatively charged puddles. This means
effectively that even at minimal conductivity the chemical potential $\mu $
locally is finite, rather than zero, albeit small on average. Physically
this implies that in addition to the novel constant contribution due to pair
creation, there is an ordinary contribution due to acceleration of electrons
like in ordinary metal. In ballistic regime it grows linearly with time.

The experiment \cite{Andrei08} shows that the amplitude of the random Fermi
energy increases linearly with temperature $\mu _{T}=\mu _{0}+\alpha T$. For
example, for the $0.5\mu m$ long sample $\mu _{0}=8meV$ and $\alpha
=0.1meV/K $. The difference between $\sigma _{\mu }$ and $\sigma _{\mu =0}$
is equal to the integral in Eq.(\ref{cond}) over the two regions around the
Dirac points determined by $\left\vert h\left( \mathbf{k}\right) \right\vert
<\mu $. That way one obtains for $t>>t_{\gamma }$

\begin{equation}
\sigma _{\mu }\left( t\right) -\sigma _{\mu =0}\left( t\right) =\frac{e^{2}}{%
h}\left[ \frac{\sqrt{3}\mu t}{\hslash }-Si\left( \frac{\sqrt{3}\mu t}{%
\hslash }\right) \right] ,  \label{sig_mu}
\end{equation}%
which is a monotonically increasing function of the product $\mu t$ only ($%
Si $ is the sine integral function). In Fig.2 we fit the value of the
ballistic effective time $t_{bal}$ $\simeq 2\cdot 10^{-13}s$, which is of
the same order of magnitude as for the $0.5\mu m$ long sample, $%
L/v_{g}\simeq 5\cdot 10^{-13}s$.

4. \textit{Discussion and summary.}

To summarize, we studied the dynamics of the particle - hole pair creation
by calculating the time evolution of current density, particle - hole number
and energy after the electric field is switched on. After a brief transient
period (of order of several $t_{\gamma }=\hslash /\gamma $) the current
density approaches a finite value. The minimal DC electric conductivity at
zero temperature is $\frac{\pi }{2}\frac{e^{2}}{h}$, different from an
accepted value $\frac{4}{\pi }\frac{e^{2}}{h}$. The later value was obtained
for nonideal systems by taking various limits (impurity strength etc.) or in
theory of finite size effects \cite{Beenaker} and does not characterize an
ideal pure infinite graphene sheet. At finite temperature $T$ the current
density diminishes on the scale of $t_{T}=\hslash /T=\frac{\gamma }{T}%
t_{\gamma }$. Therefore the phenomenon of finite resistivity without
dissipation disappears unless there exists a shorter time scale intercepting
the process like $2\pi /\omega $ for AC field, relaxation time $\tau $ for
scattering off impurities or phonons or ballistic flight time $t_{bal}$ for
finite samples. The effect of small random chemical potential was also
considered.

Let us now address the issue of the validity of the linear response
approximation used. Since the model does not provide a channel of
dissipation, the problem is nontrivial. Where does the Joule heat $\sigma
E^{2}$ go? The dynamical approach allows us to calculate the evolution of
energy as well as to go beyond linear response. Of course the energy
continuously increases with time and at certain time approaches the
conduction band edge at which stage linear response breaks down. We
calculated the evolution of current density, energy and pair number beyond
linear response and found that Bloch oscillations set in with a period of $%
t_{Bloch}=\frac{\hslash }{eaE}=\frac{\gamma }{eaE}t_{\gamma }$. The range of
applicability of the linear response was also determined. The average
current over larger times is zero. This means that at very high fields the
minimal conductivity phenomenon disappears. However in order to reach the
conditions for observation of the Bloch oscillations in graphene all other
time scales $\tau ,t_{T},t_{bal},2\pi /\omega $ should be larger than $%
t_{Bloch}$. Additional phenomena beyond linear response as well as their
relation to the Schwinger's calculation of the pair creation rate [ \cite%
{Schwinger,Gitman}] is under investigation.

\acknowledgments We are grateful to H.C. Kao, E. Kogan, E. Sonin, W.B. Jian,
E. Andrei, R. Krupke and V. Zhuravlev for discussions. Work was supported by
NSC of R.O.C. grant \#972112M009048 and MOE ATU program. M.L. acknowledges
the hospitality and support at Physics Department of NCTU.

\end{document}